\documentclass[runningheads]{cl2emult}

\usepackage{makeidx}  % allows index generation
\usepackage{graphicx} % standard LaTeX graphics tool
\usepackage{subeqnar} % subnumbers individual equations
\usepackage{multicol} % used for the two-column index
\usepackage{cropmark} % cropmarks for pages without
\usepackage{physbb}   % centering for subeqnar.sty
\makeindex            % used for the subject index
\usepackage{amssymb}
\def\greaterthansquiggle{\raise.3ex\hbox{$>$\kern-.75em\lower1ex\hbox{$\sim$}}}
\def\lessthansquiggle{\raise.3ex\hbox{$<$\kern-.75em\lower1ex\hbox{$\sim$}}}
\newcommand{\beq}{\begin{equation}}
\newcommand{\eeq}{\end{equation}}
\newcommand{\beqa}{\begin{eqnarray}}
\newcommand{\eeqa}{\end{eqnarray}}
\newcommand{\beqan}{\begin{eqnarray*}}
\newcommand{\eeqan}{\end{eqnarray*}}
\newcommand{\ba}{\begin{array}}
\newcommand{\ea}{\end{array}}
\newcommand{\no}{\nonumber}
\newcommand{\ra}{\rightarrow}

\newcommand{\ve}{\varepsilon}
\newcommand{\vp}{\varphi}

\newcommand{\Ha}{{\cal H}}

\newcommand{\R}{{\cal R}}

\def\nz{\ifmmode {I\hskip -3pt N} \else {\hbox {$I\hskip -3pt N$}}\fi}
\def\zz{\ifmmode {Z\hskip -4.8pt Z} \else
       {\hbox {$Z\hskip -4.8pt Z$}}\fi}
\def\qz{\ifmmode {Q\hskip -5.0pt\vrule height6.0pt depth 0pt
       \hskip 6pt} \else {\hbox
       {$Q\hskip -5.0pt\vrule height6.0pt depth 0pt\hskip 6pt$}}\fi}
\def\rz{\ifmmode {I\hskip -3pt R} \else {\hbox {$I\hskip -3pt R$}}\fi}
\def\cz{\ifmmode {C\hskip -4.8pt\vrule height5.8pt\hskip 6.3pt} \else
       {\hbox {$C\hskip -4.8pt\vrule height5.8pt\hskip 6.3pt$}}\fi}
\begin{document}
\title*{Generalized Bowen--York Initial Data}
\toctitle{Generalized Bowen--York Initial Data}
% allows explicit linebreak for the table of content
%
%
\titlerunning{Generalized Bowen--York Initial Data}
% allows abbreviation of title, if the full title is too long
% to fit in the running head
%
\author{R.~Beig
}
\authorrunning{R.~Beig}
% if there are more than two authors,
% please abbreviate author list for running head
%
%
\institute{Institut f\"ur Theoretische Physik, Universit\"at Wien, 
Boltzmanngasse 5, \\ A--1090 Wien, Austria}
%%\preprint{UWThPh1999-3}
\maketitle              % typesets the title of the contribution

\begin{abstract}
A class of vacuum initial-data sets is  described which are based on
certain expressions for the extrinsic curvature first studied and
employed by Bowen and York. These expressions play a role for the
momentum constraint of General Relativity which is analogous to the
role played by the Coulomb solution for the Gauss-law constraint
of electromagnetism.
\end{abstract}

\section{Introduction}
In this lecture I wish to study a specific class of solutions to the
initial-value constraints in vacuo, or, rather the `momentum' part of
these equations, which are a generalization of ones first put forward
by Bowen and York ([1],[2]). While these solutions are certainly special,
they turn out to be very useful. In particular, many of the initial-data
sets currently used by numerical relativists are Bowen--York initial data
(BY initial data), in the sense that they are based on the explicit
extrinsic curvature expressions first written down in ([1],[2]).
Although we shall in this work mainly be concerned with the momentum
constraints, the solutions we shall study can only be understood as
ingredients to solutions to the full set of initial-value constraints.

We first recall the notion of an initial-data set (IDS). This consists
of a triple $(\bar \Sigma, \bar h_{ij},\bar K_{ij})$, where $\bar \Sigma$
is a 3-manifold, $\bar h_{ij}$ a positive-definite metric on $\bar \Sigma$
and $\bar K_{ij}$ a symmetric tensor field on $\bar \Sigma$. This IDS is
called a {\bf vacuum IDS}, if the following system of equations is
satisfied
\beqa
\bar D^j(\bar K_{ij} - \bar h_{ij} \bar K) &=& 0 \\
\bar \R + \bar K^2 - \bar K^{ij} \bar K_{ij} &=& 0,
\eeqa
where $\bar D_i$ is the Levi Civita covariant derivative associated with
$\bar h_{ij}$, $\bar \R$ the scalar curvature and $\bar K := \bar h^{ij}
\bar K_{ij}$. Given a vacuum IDS, there is a spacetime $M$ with Ricci flat
Lorentz metric $g_{\mu\nu}$, in which $\bar \Sigma$ is a Cauchy surface and
$\bar h_{ij}$ (resp. $\bar K_{ij}$) are the intrinsic metric (resp.
extrinsic curvature) induced on $\bar \Sigma$.

We now review the `conformal method' for solving the vacuum initial-value
constraints. This yields solutions $(\bar h_{ij},\bar K_{ij})$ of
Equ.'s (1,2) for which $\bar K \equiv$ const. Then, if we write
\beq
\bar K_{ij} = \bar \kappa_{ij} + \frac{1}{3} \bar h_{ij} \bar K,
\eeq
it follows that
\beq
\bar D^j \bar \kappa_{ij} = 0 \qquad \mbox{and} \qquad
\bar \kappa = \bar  \kappa_{ij} \bar h^{ij} = 0.
\eeq
Thus $\bar \kappa_{ij}$ is a ``transverse-tracefree'' (TT-)tensor. The
conformal method rests on two identities.
\begin{description}
\item[Fact a):] Let $h_{ij}$ and $\bar h_{ij}$ be two conformally related
metrics, i.e.
\beq
\bar h_{ij} = \phi^4 h_{ij}, \qquad \phi > 0.
\eeq
Then, defining the conformal Laplacian $L_h$ to be
\beq
L_h := - h^{ij} D_i D_j + \frac{1}{8} \R[h] = - \Delta + \frac{1}{8} \R
\eeq
we have that
\beq
L_{\bar h}(\phi^{-1} \psi) = \phi^{-5} L_h \psi, \qquad \phi > 0.
\eeq
Setting in (7) $\phi = \psi$, it follows that
\beq
L_{\bar h} 1 = \frac{1}{8} \R[\bar h] = \phi^{-5} L_h \phi .
\eeq
\item[Fact b):] Suppose $K_{ij} = K_{(ij)}$ is trace-free. Then
\beq
\bar D^j \bar K_{ij} = \phi^{-6} D^j K_{ij}, \qquad \phi > 0,
\eeq
where $\bar K_{ij} = \phi^{-2} K_{ij}$. Combining these facts we can make
the following observation: Suppose $K_{ij}$ is a TT-tensor with respect
to the metric $h_{ij}$. Then, for any $\phi > 0$ and any constant $\bar K$,
\beq
\bar K_{ij} = \phi^{-2} K_{ij} + \frac{\bar K}{3} \phi^2 h_{ij}
\eeq
satisfies Equ. (1). Furthermore, if $\phi$ satisfies
\beq
L_h \phi = \frac{1}{8} K_{ij} K^{ij} \phi^{-7} - \frac{1}{12} \bar K^2 \phi^5,
\eeq
then $\bar h_{ij} = \phi^4 h_{ij}$ satisfies Equ. (2). Thus, solving the
constraints (1,2) amounts to choosing a `background metric' $h_{ij}$,
finding a TT-tensor $K_{ij}$ with respect to $h_{ij}$ and solving Equ. (11)
for given $(h_{ij},K_{ij})$ and a choice of constant $\bar K$.
\end{description}

We will need global conditions in order for the above program to go through.
The two cases of greatest interest is, firstly, the case where $\bar \Sigma$
is compact (``cosmological case''), and here it is natural to take the
background fields $(h_{ij},K_{ij})$ to be defined on $\Sigma = \bar \Sigma$.
Depending on the global conformal nature of $h_{ij}$, the sign of
$\bar K$ and on whether $  \bar \kappa_{ij}$ is zero or non-zero, there is
an
exhaustive list of possibilities [3] for which Equ. (11) can be solved.

The other case is that where $(\bar \Sigma, \bar h_{ij},\bar K_{ij})$ is
asymptotically flat. This means that there is a compact subset
$K \subset \bar \Sigma$, so that $\bar \Sigma \setminus K$ consists of a
finite number of asymptotic ends. An asymptotic end is a set diffeomorphic
to ${\bf R}^3 \setminus B$, where $B$ is a closed ball. Furthermore, in
the coordinate chart given by this diffeomorphism, $\bar h_{ij}$ should
satisfy
\beq
\bar h_{ij} - \delta_{ij} = O \left(\frac{1}{\bar r}\right), \qquad
\bar r^2 = \bar x^i \bar x^j \delta_{ij} ,
\eeq
\beq
\bar K_{ij} = O \left( \frac{1}{\bar r^2}\right)
\eeq
and $\partial \bar h_{ij} = o(1/\bar r^2), \ldots, \partial \bar K_{ij} =
O(1/\bar r^3), \ldots$ for a few derivatives. Since we require
$\bar K =$~const, it follows that $\bar K = 0$.

Now the problem of solving the constraints takes the following form: One
first picks an asymptotically flat 3-metric $h_{ij}$ on $\bar \Sigma$ and
a TT-tensor $K_{ij}$ on $(\bar \Sigma,h_{ij})$. One then has to solve
Equ. (11) with $\bar K = 0$ and the boundary condition that $\phi \ra 1$ 
at infinity.

A very convenient alternative is to take $(h_{ij},K_{ij})$ to be defined
on a compact manifold $\Sigma$, the many-point compactification of
$\bar \Sigma$. The ``infinities'' of $\bar \Sigma$ are now replaced by a 
finite number of points $\Lambda_\alpha \in \Sigma$, which we call
punctures. The Equ. (11) should then be replaced by
\beq
L_h \phi = \frac{1}{8} K_{ij} K^{ij} \phi^{-7} + 4\pi(c_{1}\delta_1 +
c_{2}\delta_2 + \ldots),
\eeq
where $\delta_{\alpha}$ is the delta distribution supported at
$\Lambda_\alpha$
and $c_\alpha$ are positive constants.
Note that $\bar K = 0$. Let $h_{ij}$ be a smooth metric on $\Sigma$ and 
$K_{ij}$ be smooth on $\Sigma$ except, perhaps, at the points $\Sigma_\alpha$,
where it may blow up like $1/r^4$, where $r^2 = \delta_{ij} x^i x^j$ with
$x^j$ being Riemann normal coordinates centered at $\Lambda_\alpha$.
Then a solution $\phi$ of Equ. (14) behaves like $1/r$ near 
$\Lambda_\alpha$. If a positive global solution of Equ. (14) exists, then
$\bar h_{ij} = \phi^4 h_{ij}$, $\bar K_{ij} = \phi^{-2} K_{ij}$ satisfy
the conditions of asymptotic flatness in the ``inverted'' coordinates
$\bar x^i = x^i/r^2$. For the conditions under which $\phi > 0$ exists,
see e.g. [4].

\section{The Gauss\ law constraint of electromagnetism}
Before turning to the methods for obtaining TT-tensors, it is very
instructive to use, by means of analogy, the equation
\beq
\mbox{div }E = D^i E_i = 0
\eeq
on a Riemann 3-manifold $(\Sigma,h_{ij})$, that-is-to-say the Gauss\
constraint of electrodynamics. This, like the TT-condition, is an
{\bf underdetermined elliptic system}. This means that the ``symbol map'',
namely the linear map $\sigma(k)$, sending
\beq
e_i \in {\bf R}^3 \mapsto k^i e_i \in {\bf R} \qquad
(k \in {\bf R}^3, k \neq 0)
\eeq
is onto. (The symbol map is essentially the Fourier transform of the
highest-derivative term of a partial differential operator.) 
Note, first of all, that, when $\bar h_{ij} = \phi^4 h_{ij}$
$(\phi > 0)$, $\bar E_i = \phi^{-2} E_i$, we have that
\beq
\bar D^i \bar E_i = \phi^{-6} D^i E_i .
\eeq
We will thus again take $\Sigma$ to be compact, imagining that we are
either in the cosmological case or that $\Sigma$ is the conformally
compactified, asymptotically flat 3-space $\bar \Sigma$. Our first point 
is to show two different ways of solving (15). To describe the first
one, which is the analogue of the ``York method'' for solving the
momentum constraint, we observe that the gradient operator grad, sending
$C^\infty$-functions on $\Sigma$ to $\Lambda^1(\Sigma)$, the 1-forms on 
$\Sigma$, is (minus) the formal adjoint of div under the natural 
$L^2$-inner product on $(\Sigma,h_{ij})$. It is thus, on general
grounds (see Appendix on Sobolev Spaces and Elliptic Operators in [5])
true that there is a direct-sum decomposition of
$\Lambda^1(\Sigma)$ as
\beq
\Lambda^1(\Sigma) = \mbox{grad }(C^\infty(\Sigma)) \oplus \ker
\mbox{ div},
\eeq
and this decomposition is orthogonal in the  $L^2$-sense.
Furthermore the second summand in (18) is 
infinite-dimensional. The relation (18) tells us how to find the elements
of ker~div, i.e. the solutions of Equ. (15). Namely, pick an arbitrary
1-form $\omega_i$ and write
\beq
\omega_i = D_i \vp + E_i ,
\eeq
where $E_i \in \ker \mbox{ div}$. Hence
\beq
D^i \omega_i = \mbox{div grad }\vp = \Delta \vp.
\eeq
Since the left-hand side of (20) is a divergence, it is orthogonal to
$\ker \Delta$. Thus $\Delta^{-1}\mbox{ div }\omega$ exists. Consequently,
$E_i$ can be written as 
\beq
E = [{\bf 1} - \mbox{grad (div grad})^{-1} \mbox{ div}]\omega ,
\eeq
and this is the general solution of (15). The Equ. (21) is a special case
of the way to solve an underdetermined elliptic system. It has the feature 
that it is non-local, since it involves the operation of taking the inverse
of div~grad.

There is another way of solving Equ. (15). Observe that any 1-form $E_i$
of the form ($\ve_{ijk}$ is the volume element on $\Sigma$)
\beq
E_i = \ve_i{}^{jk} D_j \mu_k ,
\eeq
or
$$
E = \mbox{rot } \mu
$$
solves Equ. (15). Is this the general solution? The answer is: yes, up to 
at most a finite-dimensional space, namely the {\bf harmonic} 1-forms, 
i.e. the elements of $\Lambda^1(\Sigma)$ which are annihilated by
$\Delta_H = (\mbox{rot})^2 - \mbox{grad div}$. More specifically we have
\beq
\ker \mbox{ div} = \ker \Delta_H \oplus \mbox{ rot }(\Lambda^1(\Sigma)).
\eeq
Note that $\ker \Delta_H = \ker \mbox{ rot } \cap \ker \mbox{ div}$.
The relation (23) gives a refinement of the decomposition (20).
Stated in a more fancy way, the decomposition (23) expresses the fact
that the de~Rham cohomology group $H^2 = \ker \mbox{ div}/\mbox{rot }
\Lambda^1(\Sigma)$ is isomorphic to $\ker \Delta_H$. Let us for simplicity
assume that $\ker \Delta_H$ is trivial. In Wheeler's words [6] we assume
that there is no ``charge without charge''. This will for example be the case
when $(\Sigma,h_{ij})$ is the standard 3-sphere. More generally, this is
the case when $(\Sigma,h_{ij})$ is of constant positive curvature
(Exercise: prove this). This implies the following: Let $E_i$ be an
arbitrary solution of Equ. (15), and $S \subset \Sigma$ be an arbitrary
embedded 2-sphere. Then the integral $\int_S E_i dS^i$ is zero, i.e.
\beq
\int_S E_i dS^i = 0.
\eeq
Suppose now that $\Sigma$ contains some punctures, at which we are willing 
to allow $D^i E_i$ to become singular. Then Equ. (24) will in general
no longer be valid, and $E = \mbox{ rot }\mu$ will no longer be the
general solution to Equ. (15). Suppose for simplicity that $\Sigma$ is
diffeomorphic to the 3-sphere. If there is just one puncture $\Lambda_1$
(not to be confused with $\Lambda^1(\Sigma)$!), then (24) is clearly
still valid, whether $S$ endores $\Lambda_1$ or it doesn't. But suppose
there are two punctures, $\Lambda_1$ and $\Lambda_2$. Then, either $S$
does not enclose either of them in which case (24) is valid or it does,
in which case (24) can not be expected to hold in general. The value
\beq
Q = \int_S E_i dS^i ,
\eeq
in the latter case, constitutes a 1-parameter class of obstructions to the
existence of $\mu_i$ such that $E = \mbox{ rot }\mu$. (Of course, $Q$
depends only on the homology class of $S$, not $S$ itself.) It is thus
desirable to split the general $E$ with div~$E = 0$ as a sum of one of the
form $E = \mbox{ rot }\mu$ and a set of fields parametrized by $Q$.
Imagine this latter set of fields to be distributions which are orthogonal
to fields in the first summand. Thus these will be of the form 
$E = \mbox{ grad }\vp$, and
\beq
\mbox{div grad }\vp = \rho,
\eeq
where the distribution $\rho$ is supported on $\Lambda_1 \cup \Lambda_2$
and has to satisfy
\beq
\int_\Sigma \rho dV = 0 .
\eeq
Thus we set
\beq
\rho = Q(\delta_1 - \delta_2),
\eeq
where $\delta$ is the Dirac delta distribution. If $\vp$ is the solution 
(unique up to addition of a constant) of Equ. (27), $E = \mbox{grad }\vp$
and the normal to $S$ points towards $\Lambda_2$, then Equ. (25) is
fulfilled.

Let us be more specific. When $(\Sigma,h_{ij})$ is the standard 3-sphere, 
it can be imagined to be the 1-point conformal compactification (conformal
compactification by inverse stereographic projection from the origin, say)
of flat ${\bf R}^3 = \bar \Sigma$ (which in turn can be viewed to be a
standard $t =$~const hyperplane of the Minkowski spacetime of Special
Relativity). Suppose the origin of $\bar \Sigma$ corresponds to $\Lambda_1$
and $\Lambda_2$ is the antipode of $\Lambda_1$ on $S^3$. Thus $\Lambda_1$
should be viewed as a point where the field becomes singular and $\Lambda_2$
as the point-at-infinity. (In the gravity case, of course, singularities
do not make good sense, sp all punctures will play the role of
points-at-infinity. That this is possible, is due to the fact that
$\bar h_{ij}$ on $\bar \Sigma$ is not a fixed element of the theory, but
$\bar h_{ij} = \phi^4 h_{ij}$, and $\phi$ satisfies Equ. (14).) We now
undo the stereographic projection and write
\beq
\bar E_i = \phi^{-2} D_i \vp ,
\eeq
where $\vp$ solves (26) and $\phi$ solves
\beq
L_h \phi = 4 \pi \delta_1 .
\eeq
The resulting field $\bar E_i$ is nothing but the Coulomb field on 
$\bar \Sigma = {\bf R}^3$
with charge $Q$ sitting at the origin.

The above situation can be slightly generalized. Let $\rho$ be an arbitrary
distribution on $\Sigma$ such that
\beq
\int_\Sigma \rho dV = 0.
\eeq
In particular $\rho$ could be a smooth function. Then, if there are no
harmonic 1-forms, the general solution of
\beq
D^i E_i = \rho
\eeq
is of the form
\beq
E = \mbox{rot }\mu + \mbox{grad } \vp ,
\eeq
where
\beq
\mbox{div grad }\vp = \rho.
\eeq
The second term in Equ. (33) could be called the generalized Coulomb
field corresponding to the source $\rho$ or, alternatively, the
``longitudinal'' solution of Equ. (32).

\section{The momentum constraint}
We will now turn to the gravity case, i.e. the momentum constraints.
The solutions originally due to Bowen and York (resp. our generalization
thereof) will turn out to be close analogues of the Coulomb field (resp.
the generalized Coulomb field), as described above. When $(\Sigma,h_{ij})$
is again taken to be standard $S^3$ with antipodal punctures $\Lambda_1$,
$\Lambda_2$ there will be a 10-parameter set of sources for the
TT-condition $D^j K_{ij} = 0$, $K = 0$, so that, under stereographic
projection relative to $\Lambda_1$, the TT-tensor on punctured ${\bf R}^3$
corresponding to the longitudinal solution of the inhomogeneous
TT-condition on $\Sigma$ contain exactly the ones written down by Bowen
and York. How does the number 10 enter here? In the Maxwell case we had
1, which was the null space of grad, which in turn is the adjoint of the
operator div. In the gravity case 10 arises as the null space of the adjoint to the
operator div, acting on symmetric, trace-free tensors, namely the conformal
Killing operator. This null space, in turn, is the space of conformal 
Killing vectors, which has at most 10 dimensions on a 3-manifold $\Sigma$.
We now have to explain these things in more detail. We again take
$(\Sigma,h_{ij})$ to be compact. The underdetermined elliptic system
\beq
(\mbox{div }K)_i = D^j K_{ij} = 0, \qquad K = 0
\eeq
is, in complete analogy with the Maxwell case, solved by the ``York
decomposition''
\beq
Q_{ij} = (LW)_{ij} + K_{ij},
\eeq
where $Q_{ij}$ is an arbitrary symmetric, trace-free tensor on $(\Sigma,
h_{ij})$ and $L$ is the conformal Killing operator
\beq
(LW)_{ij} := D_i W_j + D_j W_i - \frac{1}{3} h_{ij} D^\ell W_\ell ,
\eeq
(not to be confused with the conformal Laplacian $L_h$)
which is (-1/2 times) the adjoint of div. Hence, in order for $K_{ij}$
to be in the null space of div, we have that
\beq
\mbox{div } \circ L W = \mbox{div } Q .
\eeq
The operator on the left in Equ. (38) is elliptic with null space
consisting of conformal Killing vectors (Proof: Contract with a covector
$\lambda$ and
integrate by parts!). Thus, in complete analogy to Equ. (21) in the
electromagnetic case (15), the general solution $K_{ij}$ of the
TT-condition, Equ. (35), is given by
\beq
K = [{\bf 1} - L(\mbox{div } \circ L)^{-1}\mbox{div}]Q ,
\eeq
where $Q_{ij}$ is an arbitrary symmetric, trace-free tensor.
The above procedure works for an arbitrary $(\Sigma,h_{ij})$ with $\Sigma$
compact. It would also work if $\Sigma$ was asymptotically flat (see e.g.
[7]). We now again ask the question as to whether there is a more
explicit method for finding TT-tensors where (35) is solved ``by
differentiation'' rather than ``by integration'', as in Equ. (39). There
is a positive answer, but only in the case where $(\Sigma,h_{ij})$ is
(locally) conformally flat. Luckily, this comprises many of the IDS's
which are currently in use by the numerical relativists, as we shall
describe in Sect.~5. The necessary and sufficient condition [8] for
$h_{ij}$ to be conformally flat is that the Cotton--York tensor
$\Ha_{ij}$ defined by
\beq
\Ha_{ij} = \ve_{k\ell (i} D^k \R^\ell{}_{j)}
\eeq
be zero. Note that $\Ha_{ij}$ is always symmetric, trace-free. Also, as
a consequence of the Bianchi identities, it is divergence-free. Thus
$\Ha_{ij}$ is a TT-tensor with respect to $h_{ij}$. If $h_{ij}$ is such
that $\Ha_{ij}$ vanishes, it follows that the operator $H_{ij}$,
obtained by linearization of $\Ha_{ij}$ at $h_{ij}$, is a (third-order)
partial differential operator mapping symmetric tensors $\ell_{ij}$
(we take them also to be trace-free) into tensors which are TT with
respect to $h_{ij}$. We refrain here from writing down the operator
$H_{ij}$ explicitly (see [9]). There now arises the question of whether
$K_{ij}$ given by $K_{ij} = H_{ij}(\ell)$ is the general TT-tensor. The
answer, given in [9], is again ``yes'' up to a finite dimensional set
of ``harmonic TT-tensors'', i.e. symmetric, trace-free tensors $\ell_{ij}$
which satisfy both $D^j \ell_{ij} = 0$ and $H_{ij}(\ell) = 0$. What is
the condition for the absence of such harmonic TT-tensors? In allusion to
Wheeler [6] we might describe this situation by the absence of ``momentum
without momentum''. It is shown
in [9], that this condition is exactly that, for any 2-surface $S$ 
embedded in $\Sigma$ and any conformal Killing vector (CKV) $\xi^i$ on
$(\Sigma,h_{ij})$ which is defined near $S$, there holds
\beq
\int_S K_{ij} \xi^i dS^j = 0
\eeq
for all TT-tensors $K_{ij}$. We now come to the inhomogeneous equation
\beq
D^\ell K_{i\ell} = j_i
\eeq
which is the analogue of the Maxwell equation (32). It would be nice to
have a very clear geometrical motivation for our expression for $j_i$,
but we have to leave that for future work. The proposal is that $j_i$
depends on a ``charge density $\rho$'' and a CKV $\eta^i$, as follows:
\beq
j_i(\eta) = - D^\ell(D_{[\ell} \eta_{i]} \rho) + \frac{2}{3}
(D_i D_\ell \eta^\ell)\rho + \frac{2}{3} D_i D_\ell(\eta^\ell \rho) +
\frac{2}{9} D_i((D_\ell \eta^\ell)\rho) + 4 L_{i\ell} \eta^\ell \rho,
\eeq
where $L_{ij} = \R_{ij} - \frac{1}{4} h_{ij} \R$. It is important that
Equ. (42) behaves naturally under conformal rescalings of the metric, i.e.
$\bar h_{ij} = \phi^4 h_{ij}$. This is the case since one can show that
\beq
\bar \jmath_i(\eta) = \omega^{-6} j_i(\eta).
\eeq
Thus, with $\bar K_{ij} = \omega^{-2} K_{ij}$, $\bar K_{ij}$ satisfies
$\bar D^\ell \bar K_{i\ell} = \bar \jmath_i$ provided $K_{ij}$ solves
Equ. (42).

Suppose we have an open region $\Omega \subset \Sigma$, bounded by $S$. Then
it follows from (43) that
\beq
\int_S K_{ij} \lambda^i dS^j = \int_\Omega X(\lambda,\eta) \rho dV,
\eeq
where
\beq
X(\lambda,\eta) = D^{[i} \lambda^{j]} D_{[i} \eta_{j]} + \frac{2}{3}
[\lambda^i D_i D_j \eta^j + \eta^i D_i D_j \lambda^j] - \frac{2}{9}
(D_i \lambda^i)(D_j \eta^j) + 4 L_{ij} \lambda^i \eta^j .
\eeq
The bilinear form $X$ has an important geometrical meaning. Recall that
the CKV's span a vector space {\bf W} of at most ten dimensions. (The
dimension ten is reached when $\Sigma$ is simply connected, in which
case the general CKV $\eta^i$ can be characterized by the values of
$\eta^i$ at some point $p \in \Sigma$, together with those of
$D_{[i} \eta_{j]}$, $D_j \eta^j$ and $D_i D_j \eta^j$, the so-called
``conformal Killing data'' [10].) The Lie commutator of vector fields on
$\Sigma$ induces a Lie algebra structure on {\bf W}. The form 
$X(\lambda,\eta)$, for $\eta$ and $\lambda$ {\bf both} CKV's, is nothing
but (1/3) times the Killing metric (see App. B of [11]). One can check
by explicit computation that
\beq
X(\lambda,\eta) = \mbox{constant on } \Sigma ,
\eeq
when $\lambda$ and $\eta$ are both CKV's. Thus, from (45)
\beq
\int_S K_{ij} \lambda^i dS^j = X(\lambda,\eta) \int_\Omega \rho dV .
\eeq
So, similar to the Maxwell case, $\rho$ can not in general be a 
distribution concentrated at a single point. Rather, when 
$\Omega = \Sigma$ it follows that
\beq
X(\lambda,\eta) \int_\Sigma \rho dV = 0 .
\eeq
Equ. (50) is the necessary and sufficient condition in order for Equ. (42),
with $j_i$ given by (43), to have solutions. Note that, when dim {\bf W}
is 10, the Killing metric is non-degenerate.
The solution of the equation
\beq
D^\ell K_{i\ell} = j_i(\eta)
\eeq
becomes unique when we require $K_{ij}$ to be longitudinal, i.e.
\beq
K_{ij} = (LW)_{ij} .
\eeq
It remains to solve the elliptic equation
\beq
\mbox{div }\circ LW = j(\eta) .
\eeq
The most important case is again where $(\Sigma,h_{ij})$ is the standard 
three-sphere and
\beq
\rho = - 2 \pi (\delta_1 - \delta_2) .
\eeq
To write down explicitly the 10-parameter set of solutions, it is convenient
to send $\Lambda_2$ to infinity by a stereographic projection. Then we
have again Equ. (49), but on $({\bf R}^3,\delta_{ij})$, punctured at the
origin. The CKV's on ${\bf R}^3$ fall into the following classes
\beqa
^1\eta^i(x) &=& Q^i, \qquad Q^i = \mbox{const} \\
^2\eta^i(x) &=& \ve^i{}_{jk} S^j x^k, \qquad S^i = \mbox{const} \\
^3\eta^i(x) &=& Cx^i, \qquad C = \mbox{const} \\
^4\eta^i(x) &=& (x,x)P^i - 2(x,P)x^i, \qquad P^i = \mbox{const.} 
\eeqa
Here $x^i$ are cartesian coordinates. We find
\beqa
^1K_{ij}(x) &=& \frac{3}{2r^2} [P_i n_j + P_j n_i - (\delta_{ij} - n_i n_j)
(P,n)] \\
^2K_{ij}(x) &=& \frac{6}{r^3} \ve_{k\ell(i} S^k n^\ell n_j) \\
^3K_{ij}(x) &=& \frac{C}{r^3} (3n_i n_j - \delta_{ij}) \\
^4K_{ij}(x) &=& \frac{3}{2r^4} [-Q_i n_j - Q_j n_i - (\delta_{ij} -
5n_i n_j)(Q,n)].
\eeqa
Here $n_i = x_i/r$. The constants $P_i$ in (58) play the
role of the linear ADM-momentum at $r = \infty$ and the $S^i$ in (59) are
the ADM angular momentum. Thus $P_i$ and $S_i$ are conserved under time
evolution. If we had sent the point $\Lambda_1$ to infinity, or -- what is
the same -- if we made an inversion of the form
$\bar x^i = x^i/a^2 r^2$
 (``Kelvin transform''), the $Q_i$ would be $a^2$ times the linear momentum
at $r = 0$, viewed as another infinity. Similarly (54) goes over into (57)
with $P^i = Q^i$ under Kelvin transform. The role of the constant $C$ is less
clear. It was used in [4], to construct IDS's which have future-trapped
surfaces. The constant $C$ does not correspond to a conserved quantity.

If $K_{ij}$ was a sum of the expression in (58) and that in (59) the
physical interpretation is that they characterize a single black hole with
momentum $P^i$ and spin $S^i$.

We now come to the technical result of this section. Suppose
$(\Sigma,h_{ij})$ is of constant curvature, i.e.
\beq
\R_{ijk\ell} = \frac{\R}{3} h_{k[i} h_{j]\ell}, \qquad \R = \mbox{const.}
\eeq
Suppose, further, we know a function (distribution) $G$ satisfying
\beq
\Delta G = \rho .
\eeq
In the case when $\Sigma$ is compact, this will exist provided that
$\int_\Sigma \rho dV = 0$. Then there is an explicit expression for $W_i$
solving Equ. (52), namely
\beqa
W_i &=& \frac{1}{2} \eta^j D_j D_i G + \frac{3}{2}(D_i \eta_j)D^j G +
(D_j \eta^j)D_i G + \frac{\R}{3} \eta_i G \no \\
&=& - D^j [(D_{[j} \eta_{i]}) G] - \frac{1}{6} D_i[(D_j \eta^j)G] +
\frac{1}{2} D_i D_j(\eta^j G) .
\eeqa
For an outline of the proof, see [12]. It is easy to check that, when
$\Sigma$ is flat ${\bf R}^3$, $\rho = -2\pi \delta(x)$ and $\xi^i$ runs
through (54--57), the $K_{ij}$'s given by $(LW)_{ij}$ with $W_i$ as in
Equ. (64), reduce to (58--61).

\section{Boosting a single black hole}
In this section we present a simple perturbative calculation which should
serve as a check whether the ``boost-type'' extrinsic curvature (58) gives
a sensible result for the full IDS. We assume that $(\Sigma,h_{ij})$ 
is a standard three-sphere and we try to solve the Lichnerowicz equation
(14) with $\Lambda_1$ and $\Lambda_2$ being south and north pole,
respectively. The $K_{ij}$ in (14) should be the one turning into (58)
after stereographic projection. When $K_{ij}$ vanishes and the 
three-sphere has radius $1/m$, the unique solution to (14) gives rise to
the Schwarzschild solution of mass $m$. After stereographic projection the
three-sphere punctured at $\Lambda_1$ and $\Lambda_2$ becomes flat
${\bf R}^3$, punctured at the origin. The Lichnerowicz conformal factor 
$\phi$ such that $\bar h_{ij} = \phi^4 \delta_{ij}$ for Schwarzschild is,
as is well-known,
\beq
\phi = 1 + \frac{m}{2r} .
\eeq
By the conformal invariance of the York procedure we can of course also
start from flat, punctured ${\bf R}^3$. We are thus trying to solve
\beq
\Delta \phi = - \frac{1}{8} K_{ij} K^{ij} \phi^{-7},
\eeq
with $K_{ij}$ given by Equ. (58), and $\phi$ should go to one at infinity
and have a $1/r$-singularity
near $r = 0$ so that $\phi^4 \delta_{ij}$ becomes asymptotically flat both
near infinity and near $r = 0$. We have that
\beq
K_{ij} K^{ij} = \frac{9}{2r^4} [P_i P^i + 2(P_i n^i)^2] .
\eeq
For $\phi$ we make the ansatz
\beq
\phi = 1 + \frac{m}{2r} + \psi ,
\eeq
where $\psi$ should vanish at $r = \infty$ and be regular near $r = 0$. We
only keep terms quadratic in $P_i$. It follows that
\beq
\Delta \psi = - \frac{9r^3}{16} \frac{P^2 + 2(P,n)^2}
{(r + m/2)^7}.
\eeq
Thus
\beq
\psi(x) = \frac{9}{4\pi \cdot 16} \int_{{\bf R}^3} \frac{r'{}^3}{|x-x'|}
\frac{P^2 + 2(P,n')^2}{(r' + m/2)^7} dx' .
\eeq
It is not difficult to find that
\beq
\lim_{x \ra 0} \psi(x) = \frac{P^2}{8m^2}
\eeq
and
\beq
\psi(x) = \frac{5P^2}{16 m r} + O\left( \frac{1}{r^2}\right),
\eeq
near $r = \infty$. While the constant $m$ is only a formal parameter, the
true ``observables''
are the ADM-energies at the two infinities. The ADM energy 
$M$ at $r = \infty$ is given by
\beq
M = m + \frac{5P^2}{8m} .
\eeq
The ADM energy $\bar M$ near $r = 0$ is obtained by noting that, near
$r = 0$,
\beq
d\bar s^2 = \bar h_{ij} dx^i dx^j = \left[ 1 + \frac{P^2}{8m^2} +
\frac{m}{2r} + O(r) \right]^4 \delta_{ij} dx^i dx^j .
\eeq
After inversion $\bar x^i = (m/2)^2 x^i/r^2$, this results in
\beqa
d\bar s^2 &=& \left[ 1 + \frac{P^2}{8m^2} + \frac{2\bar r}{m} +
O \left(\frac{1}{\bar r} \right)\right]^4
\left( \frac{m}{2} \right)^4 \frac{1}{\bar r^4} \delta_{ij}
d \bar x^i d \bar x^j \no \\
&=& \left[ \frac{m}{2\bar r} \left( 1 + \frac{P^2}{8m^2}\right) + 1 +
O \left(\frac{1}{\bar r^2} \right)\right] \delta_{ij}
d \bar x^i d \bar x^j .
\eeqa
Consequently,
\beq
\bar M = m + \frac{P^2}{8m} + O(P^4).
\eeq
Thus,
\beq m = \bar M - \frac{P^2}{8 \bar M} + O(P^4).
\eeq
Inserting (77) into (73) we finally obtain
\beq
M = \bar M + \frac{P^2}{2 \bar M} + O(P^4)
\eeq
or
\beq
M^2 - P^2 = \bar M^2 + O(P^4).
\eeq
Recall that with our ansatz for $K_{ij}$ an observer near $r = \infty$
``sees'' a hole at $r = 0$ with momentum $P$ and one easily finds that
an observer near $r = 0$ sees a hole at $r = \infty$ at momentum zero.
Thus Equ. (79) says that the rest-masses at both asymptotic
ends are equal.
This equality expresses the ``absence of gravitational radiation'' on the
spacetime slice $\Sigma$. If there was more than one hole, the analogue of
Equ. (79) would contain for example potential energy-contributions from
the mutual gravitational interaction between those holes, as in the
famous calculation by Brill and Lindquist [13]. See [14] for interaction
energies of more general geometries.
Had we used as our ansatz for $K_{ij}$ a sum of the one in Equ. (58) and
the one in (61) we would  obtain
\beq
M + \frac{\bar P^2}{2M} + O(\bar P^4) = \bar M + \frac{P^2}{2 \bar M} + O(P^4),
\eeq
where $\bar P$ is the ADM 3-momentum of the $r=0$-end of $\Sigma$, which is
related to $Q$ by $\bar P = 4Q/M^2$. I thank U.Kiermayr for performing
this computation.

Does, for the present IDS, the relation
\beq
M^2 - P^2 = \bar M^2 - \bar P^2
\eeq
hold exactly? In this connection we should point out that the present data
is {\bf not} the same as the one induced by the Schwarzschild spacetime of
mass $M$ on a boosted maximal slice, since the metric on such slices can not
be conformally flat (see [15]).

\section{More general initial-data sets with punctures}
Let again $(\Sigma,h_{ij})$ be a compact, conformally flat manifold and
$\eta^i$ a CKV on $\Sigma$. Then consider the equation (50), namely
\beq
D^\ell K_{i\ell} = j_i (\eta),
\eeq
with $j_i(\eta)$ given by Equ. (43). Although we assumed in our heuristic
discussion that there are no ``harmonic'' TT-tensors on $\Sigma$, this
is actually not required for (81) to make sense. The simplest case is where
$\Sigma$ is a standard three-sphere and $\rho$ is a delta function
concentrated at a finite number of points $\Lambda_\alpha \in \Sigma$
$(\alpha = 1,\ldots,N)$. We have to have
\beq
\int_\Sigma \rho dV = 0
\eeq
so that $N \geq 2$. The case $N = 2$ contains the situation discussed in 
the previous section. The case of general $N$ is for some choices of 
$\eta^i$ leads to the IDS's studied by Brandt and Br\"ugmann [16].

Another interesting case is that where $(\Sigma,h_{ij})$ is
$S^2 \times S^1(a)$, i.e. the unit-two sphere times the circle of length
$a$. This is conformally flat but not of constant curvature. The
equation
\beq
L_h \phi = 4 \pi \delta_1
\eeq
has a unique positive solution $\phi$ (since $\R > 0$, see [17]). The
manifold $(\bar \Sigma = \Sigma \setminus \Lambda_1,\bar h_{ij})$ with
$\bar h_{ij} = \phi^4 h_{ij}$ is nothing but the (time-symmetric) Misner
wormhole [18]. Taking two punctures $\Lambda_1$ and $\Lambda_2$ on
$\Sigma$ at the same location in $S^2$ and at opposite points in
$S^1(a)$ and solving
\beq
L_h \phi = 4 \pi(c_{1}\delta_1 + c_{2}\delta_2),
\eeq
(which can be done by linearly superposing two solutions like that in
(84)) we get for $(\bar \Sigma,\bar h_{ij})$ two asymptotically flat
sheets joined by two Einstein--Rosen bridges [19]. These two
time-symmetric IDS's can be turned into generalized Bowen--York ones by
first solving
\beq
D^\ell K_{i\ell} = j_i(\eta)
\eeq
for some CKV $\eta^i$ and then solving
\beq
L_h \phi = \frac{1}{8} K_{ij} K^{ij} \phi^{-7} + 4\pi \sum_\alpha
c_{\alpha}\delta_\alpha .
\eeq
We just deal with Equ. (86) here. In the first case (1 puncture) we cannot
have $\int \rho = 0$, so that $\eta$ has to be in the null space of
$X(\xi,\eta)$, where $\xi$ runs through all CKV's on $(\Sigma,h_{ij})$.
But, on this manifold, all conformal Killing vectors are in fact Killing
vectors. (This would be true
for any compact Riemannian manifold except standard $S^3$, see [20] or
also App. A of [21].) Hence the CKV's just comprise 
rotations in the $S^2$-direction and a covariant constant vector in the
$S^1$-direction. The latter, from Equ. (45), is allowed, but not the
rotations. Thus, within the method of this paper, it is possible to 
boost a Misner wormhole into the
direction connecting the two wormhole throats, but it is impossible to
spin up a Misner wormhole. 
In the Einstein--Rosen case both options are
available (see Bowen, York [22], Kulkarni et al. [23] and Bowen et al. 
[24]). It is not clear to me whether the solutions
of (81) constructed by these authors, when viewed as ones on the
handle manifold $S^2 \times S^1(a)$, are longitudinal or not.

\section*{Acknowledgement}

This work was supported by Fonds zur F\"orderung der wissenschaftlichen
Forschung in \"Osterreich, Projekt Nr. P12626-PHY. I thank U.Kiermayr for
a discussion which led to the discovery of an error in a previous version
of this work.

%INDEX%%%%%%%%%%%%%%%%%%%%%%%%%%%%%%%%%%%%%%%%%%%%%%%%%%%%%%%%%%%%%%%
\clearpage
\addcontentsline{toc}{section}{Index}
\flushbottom
\printindex
%%%%%%%%%%%%%%%%%%%%%%%%%%%%%%%%%%%%%%%%%%%%%%%%%%%%%%%%%%%%%%%%%%%%%

\end{document}